\documentclass[journal=jctcce,manuscript=article]{achemso}

\usepackage{graphicx}
\usepackage{amsmath,amssymb}
\usepackage{mathtools}
\usepackage{bm}
\usepackage{physics}
\usepackage{xcolor}

\newcommand{\CUHK}{Guangdong Basic Research Center of Excellence for Aggregate Science, School of Science and Engineering, The Chinese University of Hong Kong (Shenzhen), Shenzhen, Guangdong 518172, P. R. China}

\author{Cunxi Gong}
\affiliation{\CUHK}

\author{Zirui Sheng}
\affiliation{\CUHK}

\author{Weitang Li}
\email{liwt31@gmail.com}
\affiliation{\CUHK}

\title{Entanglement structure of the dynamical phases in the sub-Ohmic spin-boson model}

\begin{document}

\begin{abstract}
The sub-Ohmic spin-boson model exhibits three distinct dynamical regimes in its spin population dynamics, classified as coherent, incoherent, and pseudo-coherent. Whether these regimes correspond to distinct spin-bath entanglement structures remains an open question. Here we address this using tree tensor network states with projector-splitting time evolution (TTN-TDVP-PS), scanning a broad grid in the sub-Ohmic $(s, \alpha)$ plane. We find that the spin entanglement entropy $S_\mathrm{spin}(t)$ reaches a stationary plateau on a timescale shorter than the polarization relaxation, enabling construction of a stationary entropy landscape from the stationary value $S_\mathrm{stable}$. Within this scalar entropy landscape, the entropy ridge broadly follows the population-based phase boundary at small $s$, but does not reproduce the two-branch structure at large $s$. The ridge remains single-valued within the incoherent region rather than separately tracking both population-based transitions. The Bloch-sphere representation provides a geometric interpretation of this behavior. The entropy plateau corresponds to trajectories settling onto constant-radius shells, with the ridge marking the parameters of smallest stationary Bloch radius. Mode-resolved bath entanglement shows that low-frequency modes dominate the environmental entropy scale and that coherent dynamics enhance bath-mode correlations beyond direct spin--mode correlations. These results establish the stationary spin entanglement entropy as a physically informative observable that complements population-based classifications of dissipative quantum dynamics.
\end{abstract}

\section{Introduction}

The spin-boson model (SBM)~\cite{Leggett1987DynamicsDissipativeTwostate} describes a two-level quantum system linearly coupled to a bosonic environment and is one of the paradigmatic models for studying dissipative quantum dynamics.
Despite its simple form, the SBM captures a broad range of physical processes, including electron transfer reactions~\cite{Kirchberg2020ChargeTransferRedox,RudolphA.MarcusNobelPrizeChemistry}, hydrogen tunneling~\cite{Suarez1991HydrogenTunnelingCondensed}, macroscopic quantum coherence~\cite{Jordan2022ShorttimeCoherenceQubit}, quantum dissipation~\cite{Shao2014DynamicsSpinbosonModel}, and quantum information processing~\cite{LeHur2009EntanglementDecoherenceDynamics}.
Because a two-state subsystem interacting with an environment can often be mapped onto an SBM-type description, the model has become a standard platform for understanding the competition between coherent tunneling and environmental dissipation.
It also serves as an important benchmark for quantum dynamics methods, since it admits analytical results in certain limits while remaining highly nontrivial in general parameter regimes~\cite{Guan2024MpsqdMatrixProduct,Li2023EfficientQuantumSimulation,Larsson2025BenchmarkingVibrationalSpectra}.
The SBM Hamiltonian consists of a tunneling term, a bias term, a set of harmonic oscillators representing the bath, and a linear spin-bath coupling term.
For a harmonic bath linearly coupled to the two-level system, the environmental influence is fully specified by the spectral density function $J(\omega)$, which encodes both the distribution of bath frequencies and the corresponding coupling strengths.
Reliable propagation of SBM dynamics has been enabled by a wide range of numerically exact methods, including path-integral approaches such as QUAPI~\cite{Makri1995TensorPropagatorIterative,Makri1995TensorPropagatorIterativea,Otterpohl2022HiddenPhaseSpinBoson}, the hierarchy equations of motion (HEOM)~\cite{Tanimura1989TimeEvolutionQuantum,Yan2004HierarchicalApproachBased,Tanimura2020NumericallyExactApproach,Ye2016HEOMQUICKProgramAccurate,Shi2025HierarchicalEquationsMotion}, generalized quantum master equations~\cite{Shi2003NewApproachCalculating,Curchod2018InitioNonadiabaticQuantum}, non-Markovian quantum state diffusion~\cite{Strunz1999OpenSystemDynamics,Suess2014HierarchyStochasticPure}, the multi-configuration time-dependent Hartree (MCTDH) method and its multilayer extension~\cite{Meyer1990MulticonfigurationalTimedependentHartree,Wang2003MultilayerFormulationMulticonfiguration,Wang2008CoherentMotionLocalization,Wang2010CoherentMotionLocalization,Wang2015MultilayerMulticonfigurationTimeDependent}, and the numerical renormalization group and density matrix renormalization group (DMRG)~\cite{Bulla2005NumericalRenormalizationGroup,Anders2006SpinPrecessionRealtime,Anders1995EquilibriumNonequilibriumDynamics,White1992DensityMatrixFormulation,White1993DensitymatrixAlgorithmsQuantum,Baiardi2020DensityMatrixRenormalization,Ren2022TimedependentDensityMatrix,Haegeman2011TimeDependentVariationalPrinciple,Haegeman2016UnifyingTimeEvolution,Kloss2018TimedependentVariationalPrinciple,Milsted2013VariationalMatrixProduct,Ma2018TimedependentDensityMatrix}.
These methods have primarily been applied to characterize population dynamics and locate phase boundaries.

Based on the spin population dynamics $\langle\sigma_z(t)\rangle$, Otterpohl \textit{et al.}~\cite{Otterpohl2022HiddenPhaseSpinBoson} classified the sub-Ohmic SBM into coherent (damped oscillatory), incoherent (monotonic decay), and pseudo-coherent (single minimum followed by localization) regimes.
This population-based phase diagram provides a useful dynamical classification, but does not by itself determine whether these regimes correspond to distinct spin-bath entanglement structures.
This distinction is especially important in the sub-Ohmic regime, where slow bath modes produce long memory effects and make the relation between dynamical classification and entanglement structure nontrivial.

To address this question, we employ tree tensor network states (TTNS)~\cite{Li2024OptimalTreeTensor,Nakatani2013EfficientTreeTensorb,Gunst2018T3NSThreeLeggedTree} combined with the projector-splitting time-dependent variational principle (TDVP-PS)~\cite{Baiardi2020DensityMatrixRenormalization,Li2020FiniteTemperatureTDDMRGCarrier,Li2020NumericalAssessmentAccuracy,Ren2018TimeDependentDensityMatrix,Ren2022TimedependentDensityMatrix}.
TTNS generalize matrix product states (MPS)~\cite{Ostlund1995ThermodynamicLimitDensity,Schollwock2011DensitymatrixRenormalizationGroup} to hierarchical tree topologies suited for large structured environments, closely related in spirit to ML-MCTDH.
The Hamiltonian is represented as a tree tensor network operator (TTNO), constructed automatically using the bipartite-graph approach~\cite{Ren2020GeneralAutomaticMethod,Li2024OptimalTreeTensor}.
The algorithms are implemented in the open-source package Renormalizer~\cite{Ren2022TimedependentDensityMatrix}.
Within this framework, the spin entanglement entropy~\cite{Eisert2010ColloquiumAreaLaws,Gu2004EntanglementQuantumPhase}, mode-resolved bath entanglement, and spin--mode 2-body entropies are all directly accessible from the propagated wavefunction.

In this work, we scan the sub-Ohmic $(s,\alpha)$ parameter space and construct a stationary entropy landscape from the time-averaged spin entanglement entropy.
This landscape broadly follows the population-based phase boundary at small $s$ but does not reproduce the two-branch structure at larger $s$, where the entropy ridge remains single-valued and lies within the incoherent region rather than separately tracking the coherent-incoherent and incoherent-pseudo-coherent transitions.
The Bloch-sphere representation provides a geometric interpretation of this behavior.
Mode-resolved analysis further reveals that low-frequency modes dominate the environmental entanglement and that coherent spin dynamics enhance bath-mode correlations beyond direct spin--mode correlations.

The remainder of this paper is organized as follows.
Section~\ref{method} presents the theoretical framework, including the spin-boson Hamiltonian, bath discretization, tensor-network representation, TDVP time evolution, automatic TTNO construction, and entanglement measures.
Section~\ref{results} presents the numerical results: Sec.~\ref{results:sigma} constructs the population-based dynamical phase diagram; Sec.~\ref{results:spin-ent-pd} presents the stationary entropy landscape with Bloch-sphere analysis; and Sec.~\ref{results:ent} analyzes the mode-resolved bath entanglement dynamics.
Section~\ref{conclusion} summarizes the conclusions.

\section{Methodology}
\label{method}

This section presents the theoretical and numerical framework used in this work. 
We first introduce the spin-boson Hamiltonian, the bath discretization, and the initial state. 
We then define the population and entanglement observables used in the analysis. 
Finally, we describe the tree tensor network representation, the automatic TTNO construction, the TDVP-PS time evolution scheme, and the simulation workflow.

\subsection{Spin-Boson Model and Bath Discretization}
\label{method:sbm}

\subsubsection{Model Hamiltonian}

The spin-boson model describes a two-level quantum system linearly coupled to a bosonic bath. 
The Hamiltonian reads
\begin{equation}
\label{eq:spin-boson model_ham}
    \begin{aligned}
        \hat{H}_{\textrm{SB}} = &-\frac{1}{2} \hbar\Delta\hat{\sigma_x} + \frac{1}{2} \epsilon \hat{\sigma_z} +\sum_{j}\left( \frac{1}{2}m_{j} \omega_{j}\hat{x}^2_{j} + \frac{\hat{p}^2_{j}}{2m_{j}} \right)\\
        &+ \frac{1}{2} \hat{\sigma_z} \sum_{j}c_{j}\hat{x}_{j}.
    \end{aligned}
\end{equation}
The first two terms describe the isolated two-level system, where $\Delta$ is the tunneling matrix element, $\epsilon$ is the bias, and $\hat{\sigma}_x$ and $\hat{\sigma}_z$ are Pauli operators.
The third term represents the bath as a collection of harmonic oscillator modes with coordinates $\hat{x}_{j}$, momenta $\hat{p}_{j}$, masses $m_{j}$, and frequencies $\omega_{j}$.
The last term gives the linear coupling between the spin and the bath, with coupling constants $c_{j}$.

The influence of the bath is characterized by the spectral density function. 
In this work, we consider a sub-Ohmic spectral density with an exponential cutoff:
\begin{equation}
\label{eq:ohmic_sdf}
    \begin{aligned}
        J(\omega) &= \sum_j \frac{c_j^2}{2\omega_j} \delta(\omega-\omega_j) \\
        &= 2\alpha \frac{\omega^s}{\omega_c^{s-1}}  e^{-\frac{\omega}{\omega_c}},
    \end{aligned}
\end{equation}
where $\alpha$ is the dimensionless spin-bath coupling strength, $s$ is the spectral exponent, and $\omega_c$ is the cutoff frequency. 
The sub-Ohmic regime corresponds to $0<s<1$, where the low-frequency part of the bath plays an especially important role in the dissipative dynamics.

\subsubsection{Bath Discretization}

To perform wavefunction-based tensor network simulations, the continuous spectral density is discretized into $N_b$ bosonic modes. 
The finite upper cutoff of the discretized bath is accounted for by the adiabatic renormalization factor
\begin{equation}
\label{eq:delta_eff}
    \begin{aligned}
    \Delta_{\textrm{eff}}
    =
    \Delta
    \exp\left[
    - \int_{\omega_{\max}}^{\infty}
    d\omega\,
    \frac{J(\omega)}{\omega^2}
    \right],
    \end{aligned}
\end{equation}
where $\omega_{\max}$ is the largest retained discretized bath frequency.
This factor represents the contribution of bath modes above the explicit discretization window to the tunneling renormalization.
In the implementation, the same expression is evaluated in the Renormalizer spectral-density convention, which introduces the corresponding convention-dependent prefactor.
The coupling strength of each discretized mode is determined by
\begin{equation}
\label{eq:general_cj}
    c^2_j = \frac{2}{\pi}\omega_j\frac{J(\omega_j)}{\rho(\omega_j)} ,
\end{equation}
where $\rho(\omega)$ is the density of discretized bath modes. 
The mode frequencies are generated from
\begin{equation}
\label{eq:boson_density_def}
    \int_0^{\omega_j} d\omega \rho(\omega) = j, \hspace{1cm} j=1,\dots,N_b .
\end{equation}
Following the discretization strategy used for earlier SBM simulations~\cite{Wang2003MultilayerFormulationMulticonfiguration,Wang2008CoherentMotionLocalization}, we choose
\begin{equation}
\label{eq:rho_wang1}
    \rho(\omega) = \frac{N_b+1}{\omega_c}e^{-\frac{\omega}{\omega_c}} .
\end{equation}
This choice places more discrete modes in the low-frequency region, which is essential for resolving the slow bath dynamics in the sub-Ohmic regime.

\subsection{Initial State and Observables}
\label{method:observables}

The initial wave function is chosen as a Hartree product state between the spin and the bath modes,
\begin{equation}
\label{eq:initial_state}
    \ket{\Psi(0)}
    =
    \ket{\uparrow}
    \otimes
    \prod_{j=1}^{N_b}\ket{0_j},
\end{equation}
where $\ket{\uparrow}$ denotes the spin state with $\sigma_z=+1$, and $\ket{0_j}$ is the initial state of the $j$-th bosonic mode.

The primary population observable is the spin population difference
\begin{equation}
\label{eq:sigmaz_observable}
    \langle \sigma_z(t) \rangle
    =
    \bra{\Psi(t)}
    \hat{\sigma}_z
    \ket{\Psi(t)} .
\end{equation}
The dynamical classification based on $\langle\sigma_z(t)\rangle$ follows the criteria of Ref.~\cite{Otterpohl2022HiddenPhaseSpinBoson}; the corresponding phase diagram is discussed in Sec.~\ref{results:sigma}.

To analyze the spin-bath correlation structure, we compute the reduced density matrix of the spin subsystem,
\begin{equation}
\label{eq:rho_spin}
    \rho_{\mathrm{spin}}(t)
    =
    \mathrm{Tr}_{\mathrm{bath}}
    \ket{\Psi(t)}\bra{\Psi(t)} .
\end{equation}
The spin entanglement entropy is then defined as
\begin{equation}
\label{eq:S_spin}
    S_{\mathrm{spin}}(t)
    =
    -\mathrm{Tr}
    \left[
    \rho_{\mathrm{spin}}(t)
    \log \rho_{\mathrm{spin}}(t)
    \right].
\end{equation}
Since the total wave function is pure, $S_{\mathrm{spin}}(t)$ measures the entanglement between the two-level system and the full bosonic environment.

To characterize the late-time spin-bath entanglement, we define the stationary spin entanglement entropy as
\begin{equation}
\label{eq:S_stable}
    S_{\mathrm{stable}}
    =
    \frac{1}{N_{\mathrm{avg}}}
    \sum_{t_n \geq t_{\mathrm{avg}}}
    S_{\mathrm{spin}}(t_n),
\end{equation}
where $t_{\mathrm{avg}}$ is the beginning of the averaging window and $N_{\mathrm{avg}}$ is the number of time steps included in the average. 
The stationary entropy landscape is the scalar field $S_{\mathrm{stable}}(s,\alpha)$ over the $(s,\alpha)$ parameter space.
We use two one-dimensional summaries of this landscape.
Along fixed-$\alpha$ cuts, the dashed ridge in Fig.~\ref{fig:spin_ent}(e) is defined as
\begin{equation}
    s_{\mathrm{ridge}}(\alpha)
    =
    \arg\max_s S_{\mathrm{stable}}(s,\alpha).
\end{equation}
Along fixed-$s$ cuts, the peak value shown in Fig.~\ref{fig:spin_ent}(f) is
\begin{equation}
    S_{\mathrm{stable}}^{\mathrm{max}}(s)
    =
    \max_\alpha S_{\mathrm{stable}}(s,\alpha),
\end{equation}
with the corresponding $\alpha$ values used to select representative peak-entropy Bloch trajectories.

For the bath analysis, we also compute the entanglement entropy of individual bosonic modes. 
For the $j$-th mode, the one-mode reduced density matrix is
\begin{equation}
\label{eq:rho_mode}
    \rho_j(t)
    =
    \mathrm{Tr}_{\overline{j}}
    \ket{\Psi(t)}\bra{\Psi(t)},
\end{equation}
where $\overline{j}$ denotes all degrees of freedom except the $j$-th bosonic mode. 
The corresponding mode entanglement entropy is
\begin{equation}
\label{eq:S_mode}
    S_j(t)
    =
    -\mathrm{Tr}
    \left[
    \rho_j(t)\log\rho_j(t)
    \right].
\end{equation}
To form the spin--boson-mode mutual information, we compute the 2-body reduced density matrix of the spin and the $j$-th bosonic mode,
\begin{equation}
\label{eq:rho_spin_mode}
    \rho_{\mathrm{spin},j}(t)
    =
    \mathrm{Tr}_{\overline{\mathrm{spin},j}}
    \ket{\Psi(t)}\bra{\Psi(t)} ,
\end{equation}
where $\overline{\mathrm{spin},j}$ denotes all degrees of freedom except the spin and the selected mode.
The corresponding 2-body entropy is
\begin{equation}
\label{eq:S_spin_mode}
    S_{\mathrm{spin},j}(t)
    =
    -\mathrm{Tr}
    \left[
    \rho_{\mathrm{spin},j}(t)\log\rho_{\mathrm{spin},j}(t)
    \right].
\end{equation}
This entropy measures the entanglement between the selected spin--mode pair and the rest of the bath, and enters the mutual information $I_{\mathrm{spin}:j}(t)=S_{\mathrm{spin}}(t)+S_j(t)-S_{\mathrm{spin},j}(t)$.

\subsection{Tree Tensor Network Representation}
\label{method:tns}

The wave function in this work is represented by a TTNS.
It is useful to first recall MPS, which factorizes a high-dimensional wave-function coefficient tensor into a chain of local tensors.
The corresponding MPO gives an analogous tensor-network representation of many-body operators.

For the spin-boson model, however, a strictly one-dimensional chain is not always the most natural representation.
The system consists of a distinguished two-level subsystem coupled to a large set of bosonic modes, and the bath may contain correlations distributed across different frequency scales.
A TTNS generalizes the MPS chain to a hierarchical tree topology, allowing both physical nodes and purely virtual auxiliary nodes.
This additional topological flexibility is useful for representing structured spin-bath and bath-bath entanglement in large open quantum systems.

In the present simulations, the spin and discretized bosonic modes are placed on the leaf nodes of the tree, while the internal auxiliary nodes provide a hierarchical decomposition of the total wave function.
The virtual bond dimensions control the amount of entanglement retained across different bipartitions of the tree.
The Hamiltonian is represented by a TTNO, the operator counterpart of TTNS.
Applying a TTNO to a TTNS proceeds through tensor contractions on the same tree topology.

Constructing a TTNO manually is more involved than constructing an MPO because the tree topology admits more possible factorizations.
We therefore use the bipartite-graph construction, which automatically converts the sum-of-products Hamiltonian into a compact and exact TTNO representation~\cite{Ren2020GeneralAutomaticMethod,Li2024OptimalTreeTensor}.
The detailed tensor-network notation for MPS, MPO, TTNS, and TTNO can be found in Ref.~\cite{Schollwock2011DensitymatrixRenormalizationGroup,Li2024OptimalTreeTensor}.

\subsection{TDVP-PS Time Evolution}
\label{method:tdvp}

Time evolution is performed using TDVP-PS.
The exact dynamics obey the time-dependent Schr\"odinger equation,
\begin{equation}
\label{eq:tdse}
i\pdv{\ket{\Psi(t)}}{t} = \hat{H} \ket{\Psi(t)} .
\end{equation}
The Dirac--Frenkel variational principle reformulates the evolution within a variational manifold by minimizing the distance between the exact Schr\"odinger vector and the variational time derivative:
\begin{equation}
\label{eq:tdvp_basicidea}
\min \vert\vert \hat{H}\ket{\Psi(t)} - i\hbar\pdv{\ket{\Psi(t)}}{t}\vert\vert^2 .
\end{equation}
Geometrically, the TTNS wave functions form a low-dimensional manifold $\mathcal{M}$ embedded in the full Hilbert space. 
The TDVP projects the exact evolution $-i\hat{H}\ket{\Psi(t)}$ onto the tangent space of this manifold at the current state:
\begin{equation}
\label{eq:tdvp_projected}
i\frac{\partial\ket{\Psi(t)}}{\partial t}
=
\hat{P}_{\mathcal{T}_{\Psi}\mathcal{M}}
\hat{H}
\ket{\Psi(t)},
\end{equation}
where $\hat{P}_{\mathcal{T}_{\Psi}\mathcal{M}}$ is the tangent-space projector.

The projector-splitting algorithm decomposes the projected time evolution into a sequence of local tensor updates. 
Formally, the evolution over one time step can be written as
\begin{equation}
\ket{\Psi}(t_0+\delta t) =  e^{-i\hat{P}\hat{H} \delta t}\ket{\Psi}(t_0),
\end{equation}
The detailed pseudocode is provided in Ref.~\cite{Ren2022TimedependentDensityMatrix,Li2024OptimalTreeTensor}.

To validate the accuracy of the implementation, benchmark comparisons with ML-MCTDH were performed for spin population dynamics. 
The TTN-TDVP-PS results show excellent agreement with ML-MCTDH reference calculations, while offering advantages for simulations with a large number of bath modes~\cite{Li2024OptimalTreeTensor}.

\subsection{Simulation Workflow and Parameters}
\label{method:workflow}

All simulations are implemented in the open-source Renormalizer package~\cite{Ren2022TimedependentDensityMatrix}, which integrates TTN wave functions, automatic TTNO construction, TDVP time evolution, and GPU acceleration.

The workflow consists of the following steps. 
First, the continuous sub-Ohmic spectral density is discretized into $N_b$ bosonic modes. 
Second, the spin-boson Hamiltonian is constructed in TTNO form using the bipartite-graph method. 
Third, the initial Hartree-product TTNS wave function is prepared. 
Fourth, the wave function is propagated in time using TDVP-PS. 
Finally, the spin population, spin entanglement entropy, mode entanglement entropy, and spin--boson-mode mutual information are evaluated from the evolved wave function.

The main simulation parameters are summarized in Table~\ref{tab:params}. 
The parameter scan is performed over the sub-Ohmic spectral exponent $s$ and the coupling strength $\alpha$. 
The resulting trajectories of $\langle\sigma_z(t)\rangle$ are used to construct the population-based dynamical phase diagram, while the stationary spin entanglement entropy $S_{\mathrm{stable}}$ is used to construct the stationary entropy landscape.

\begin{table}[h]
\centering
\caption{Simulation parameters for the sub-Ohmic spin-boson model.}
\label{tab:params}
\begin{tabular}{lll}
\hline
Parameter & Symbol & Value \\
\hline
Tunneling element & $\Delta$ & 1 \\
Energy bias & $\epsilon$ & 0 \\
Cutoff frequency & $\omega_c$ & $10\Delta$ \\
Number of boson modes & $N_b$ & 1000 \\
Spectral exponent range & $s$ & 0.1--0.9 (grid 0.1) \\
Coupling strength range & $\alpha$ & 0.01--1.0 (grid 0.05) \\
Simulation time & $t_{\max}$ & 50 a.u. \\
Averaging start time & $t_{\mathrm{avg}}$ & 10 a.u. \\
\hline
\end{tabular}
\end{table}

\section{Numerical Results}
\label{results}

\subsection{Polarization Dynamics and Population-Based Phase Diagram}
\label{results:sigma}

We first compute the spin polarization dynamics $\langle\sigma_z(t)\rangle$ and compare the resulting dynamical classification with previous QUAPI results~\cite{Otterpohl2022HiddenPhaseSpinBoson}.
The relevant parameter space is spanned by the spectral exponent $s$ and the coupling strength $\alpha$.
To illustrate how these two parameters control the spin dynamics, we select representative points along two complementary paths in the $(s,\alpha)$ plane.

\begin{figure}
    \centering
    \includegraphics[width=\linewidth]{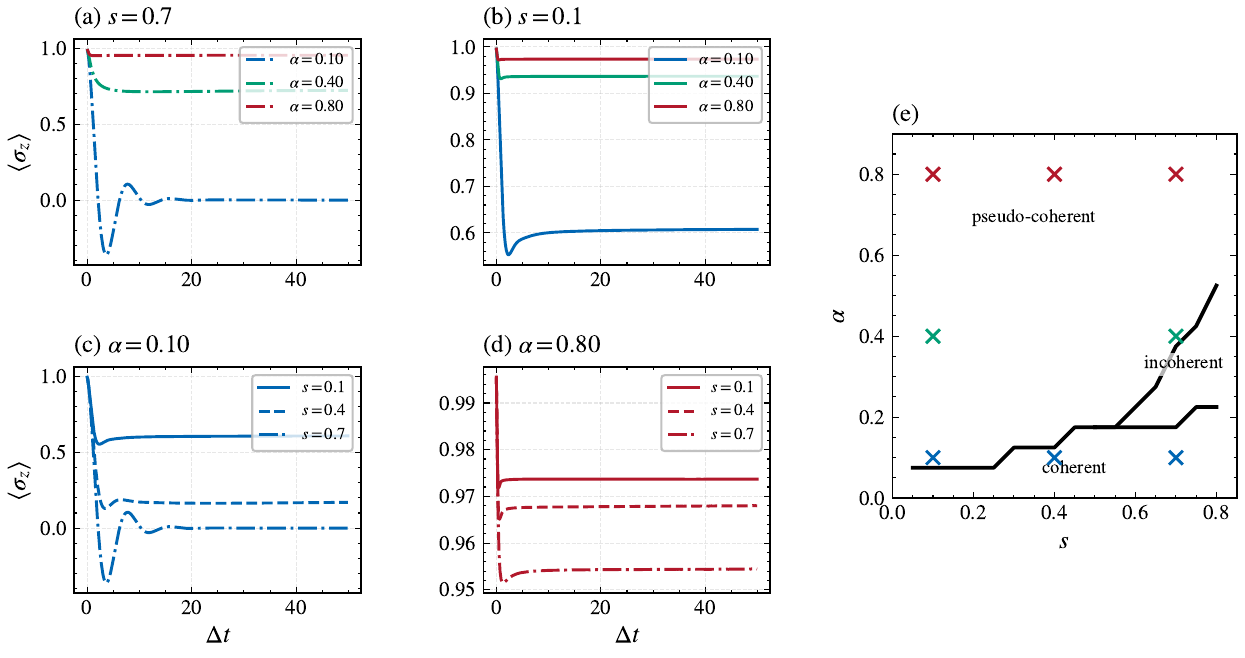}
    \caption{
    Spin polarization dynamics and population-based phase diagram of the sub-Ohmic spin-boson model.
    Panels (a) and (b) show the evolution of $\langle\sigma_z(t)\rangle$ at fixed $s$ with increasing coupling strength $\alpha$, illustrating the crossover from coherent to incoherent and pseudo-coherent dynamics.
    Panels (c) and (d) show trajectories at fixed $\alpha$ with increasing $s$, demonstrating the recovery of coherent behavior at larger spectral exponents.
    Panel (e) shows the resulting phase diagram over the $(s,\alpha)$ parameter space at $\omega_c=10\Delta$; the three regions correspond to coherent, incoherent, and pseudo-coherent spin polarization dynamics. The crosses mark the representative parameter points used in panels (a)--(d).
    }
    \label{fig:spin}
\end{figure}

In the first group [Fig.~\ref{fig:spin}(a,b)], $s$ is fixed while $\alpha$ is increased.
The dynamics evolve from coherent oscillations to incoherent decay and then to pseudo-coherent behavior.
The coherent regime is characterized by damped oscillatory motion with multiple zero-crossings, reflecting persistent quantum tunneling between the two wells.
The incoherent regime exhibits monotonic decay, indicating that environmental dissipation has suppressed coherent tunneling entirely.
The pseudo-coherent regime shows a single minimum followed by relaxation toward localization, corresponding to an initial tunneling attempt that is rapidly overdamped by the strong bath coupling.

In the second group [Fig.~\ref{fig:spin}(c,d)], $\alpha$ is fixed while $s$ is increased.
The dynamics evolve from pseudo-coherent behavior toward coherent oscillatory motion.
These two scans demonstrate that $\alpha$ and $s$ play competing roles. Increasing $\alpha$ enhances dissipation and drives localization, whereas increasing $s$ shifts spectral weight toward higher frequencies and weakens the low-frequency bath modes that are most effective at suppressing tunneling.

Following the classification of Otterpohl \textit{et al.}~\cite{Otterpohl2022HiddenPhaseSpinBoson}, we classify each trajectory into one of three regimes and perform a parameter scan with $s=0.1$--$0.9$ and $\alpha=0.01$--$1.0$ (grid spacings $\Delta s=0.1$, $\Delta\alpha=0.05$).
The resulting phase diagram [Fig.~\ref{fig:spin}(e)] agrees well with the QUAPI-based results, confirming that TTN-TDVP-PS accurately captures the dynamical phase structure.
Minor differences near phase boundaries can be attributed to our longer propagation time ($t=50$ a.u.\ vs.\ $t=10$ a.u.\ in Ref.~\cite{Otterpohl2022HiddenPhaseSpinBoson}), which can reveal slow low-frequency dynamics that do not fully develop within shorter time windows.

\subsection{Spin Entanglement Entropy and Stationary Entropy Landscape}
\label{results:spin-ent-pd}

The polarization dynamics provide a natural classification of the dynamical behavior, but they probe only the reduced motion of the spin.
To examine whether the population-defined regimes correspond to distinct spin-bath correlation structures, we compute the spin entanglement entropy $S_{\mathrm{spin}}(t)$ from the reduced density matrix of the spin subsystem.

\begin{figure}
    \centering
    \includegraphics[width=\linewidth]{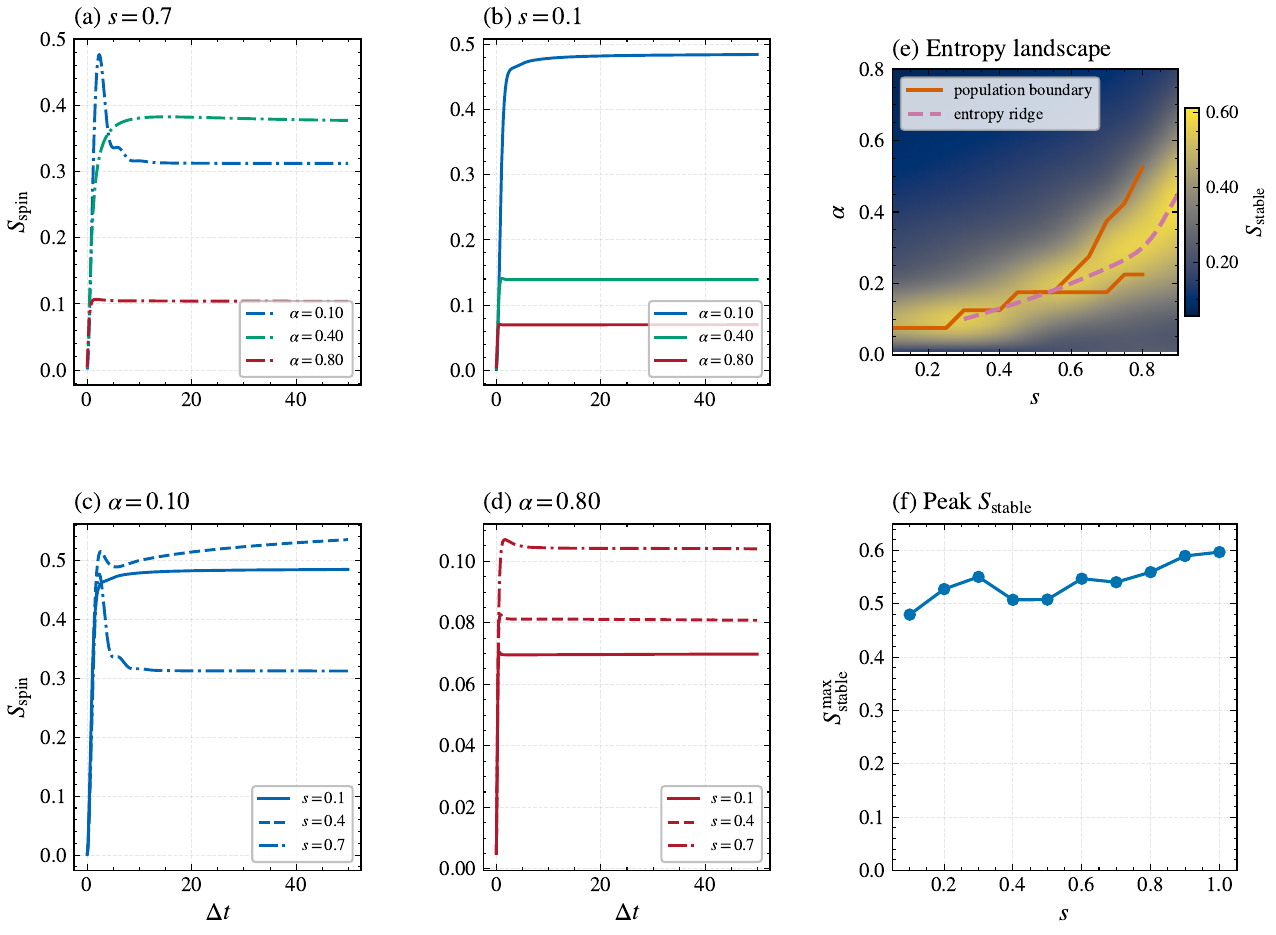}
    \caption{
    Spin entanglement entropy dynamics and stationary entropy landscape.
    Panels (a)--(d) show the spin entanglement entropy $S_{\mathrm{spin}}(t)$ for representative parameter sets, obtained from the reduced density matrix of the spin subsystem.
    Most trajectories rapidly approach a stationary plateau after an initial transient.
    Panel (e) shows the stationary entropy landscape: the color map represents the stationary spin entanglement entropy $S_{\mathrm{stable}}$ over the $(s,\alpha)$ parameter space.
    The red line denotes the population-based dynamical phase boundaries determined from $\langle\sigma_z(t)\rangle$, while the purple dashed line traces the fixed-$\alpha$ ridge $s_{\mathrm{ridge}}(\alpha)=\arg\max_s S_{\mathrm{stable}}(s,\alpha)$.
    Panel (f) shows $S_{\mathrm{stable}}^{\mathrm{max}}(s)=\max_\alpha S_{\mathrm{stable}}(s,\alpha)$, revealing a dip near $s\approx 0.4$ between two plateau regions.
    }
    \label{fig:spin_ent}
\end{figure}

Figure~\ref{fig:spin_ent}(a--d) shows $S_{\mathrm{spin}}(t)$ for the same representative points used in Fig.~\ref{fig:spin}.
For most parameter sets, the spin entanglement entropy rapidly reaches a nearly stationary plateau after an initial transient.
A notable exception is visible in Fig.~\ref{fig:spin_ent}(c) at $s=0.4$ and $\alpha=0.1$, where $S_{\mathrm{spin}}$ continues to drift upward throughout the simulation window, indicating that the spin-bath entanglement has not fully equilibrated within $t=50$ a.u. at this parameter point. This slow convergence occurs near the saddle region of the entropy landscape [Fig.~\ref{fig:spin_ent}(f)].
Extended simulations to $t=200$ a.u. confirm that the ridge-point trajectories are well converged (residual drift below 0.007), while points near the saddle and at stronger coupling continue to drift slowly (rate $\sim 10^{-4}$--$10^{-5}$/a.u.), suggesting longer equilibration timescales in that part of the parameter space.
This behavior reveals a separation of timescales, with the spin-bath entanglement reaching a plateau before the spin population dynamics becomes stationary.
Even when $\langle\sigma_z(t)\rangle$ continues to display damped oscillations beyond $t=10$ a.u., the spin-bath entanglement has already saturated.
Quantitatively, at $s=0.7$ and $\alpha=0.1$ (coherent regime), $S_{\mathrm{spin}}$ reaches 90\% of its plateau value by $t\approx 1$ a.u., while the $\langle\sigma_z\rangle$ envelope still retains significant amplitude at that time.
This implies that the late-time population oscillations occur within a state of approximately fixed spin-bath correlation, corresponding to dynamics confined to a nearly constant-entropy surface in the Bloch sphere.
We also note that in the coherent regime, $S_{\mathrm{spin}}(t)$ exhibits a transient overshoot before relaxing to its stationary plateau.
At $s=0.7$ and $\alpha=0.1$ [Fig.~\ref{fig:spin_ent}(a)], $S_{\mathrm{spin}}$ peaks at approximately 0.48 around $t\approx 2$ a.u.\ before settling to its stationary value of 0.31, an overshoot of about 53\%.
This transient excess entanglement is absent in the pseudo-coherent and incoherent regimes, where $S_{\mathrm{spin}}(t)$ approaches its plateau monotonically from below.

The plateau values show clear phase-dependent trends.
At fixed $s$, the pseudo-coherent regime has the lowest stationary spin entanglement, the incoherent regime has the highest, and the coherent regime lies in between.
For example, at $s=0.7$, the stationary values are $S_{\mathrm{stable}}\approx 0.31$ ($\alpha=0.1$, coherent), $0.38$ ($\alpha=0.4$, near the pseudo-coherent boundary), and $0.10$ ($\alpha=0.8$, deep pseudo-coherent).
The maximum at this $s$ occurs at $\alpha\approx 0.3$ ($S_{\mathrm{stable}}\approx 0.55$), within the incoherent region.
Physically, the pseudo-coherent state localizes rapidly in one well, producing an asymmetric spin-bath state with relatively low entanglement.
The incoherent regime, by contrast, reaches a reduced spin state with a smaller Bloch radius, indicating stronger spin-bath entanglement in the reduced density matrix.
The coherent regime lies between these extremes because the spin retains partial purity through its ongoing oscillatory dynamics.

Motivated by the rapid plateau formation, we define the stationary spin entanglement entropy $S_{\mathrm{stable}}$ as the time average of $S_{\mathrm{spin}}(t)$ over the stable stage (after $t=10$ a.u.; see Sec.~\ref{method:observables} for the precise definition).
Figure~\ref{fig:spin_ent}(e) shows the resulting stationary entropy landscape.
The color map represents $S_{\mathrm{stable}}$ across the $(s,\alpha)$ parameter space; the red line shows the population-based phase boundaries; and the purple dashed line traces the fixed-$\alpha$ ridge $s_{\mathrm{ridge}}(\alpha)$.
The entropy landscape forms a dome whose peak traces this ridge line.
Along fixed-$\alpha$ cuts, the ridge position shifts from $s_{\mathrm{ridge}}\approx 0.3$ at $\alpha=0.1$ to $s_{\mathrm{ridge}}\approx 0.9$ at $\alpha=0.45$, with the corresponding cut maxima remaining approximately constant ($\approx 0.55$--$0.59$) across this range.
In the strong-coupling part of the map, the peak value decreases as localization becomes dominant and even the most entangled point along a fixed-$\alpha$ cut becomes increasingly pure.

In the low-$s$ region, the entropy ridge broadly follows the population-based phase boundary within the resolution of the parameter scan.
At larger $s$, however, the two classifications differ.
The population-based phase diagram contains a branching structure separating coherent and incoherent dynamics, whereas the scalar entropy landscape does not reproduce this branching and its ridge remains a single line.
This indicates that at large $s$, the scalar entropy landscape does not separately resolve the coherent-incoherent and incoherent-pseudo-coherent transitions that appear in the population-based classification. The entropy ridge lies within the incoherent region and does not split to track both population-based boundaries.
One plausible reason is that at large $s$ the spectral density shifts weight toward higher frequencies, uniformly weakening the bath's capacity to sustain distinct entanglement structures across these two regimes.
The low-frequency modes that would otherwise differentiate the coherent and incoherent correlation structures carry less spectral weight when $s$ is large.

We note that the entropy surface also exhibits a saddle region near $s\approx 0.3$--$0.5$ at moderate $\alpha$, where $S_{\mathrm{stable}}$ dips between two local maxima.
The fixed-$\alpha$ ridge position shifts abruptly from $s_{\mathrm{ridge}}=0.3$ at $\alpha=0.1$ to $s_{\mathrm{ridge}}=0.6$ at $\alpha=0.2$, passing through this saddle.
The dip in $S_{\mathrm{stable}}^{\mathrm{max}}(s)$ near $s\approx 0.4$ [Fig.~\ref{fig:spin_ent}(f)] may be partially connected to the long equilibration timescale observed for these parameters, since $S_{\mathrm{spin}}(t)$ has not fully converged within the $t=50$ a.u.\ averaging window at the relevant coupling strengths.
The physical origin of this discontinuous shift requires further investigation, but it suggests a qualitative change in how the bath modes collectively contribute to the stationary entanglement as the coupling strength increases.

\begin{figure}
    \centering
    \includegraphics[width=0.72\linewidth]{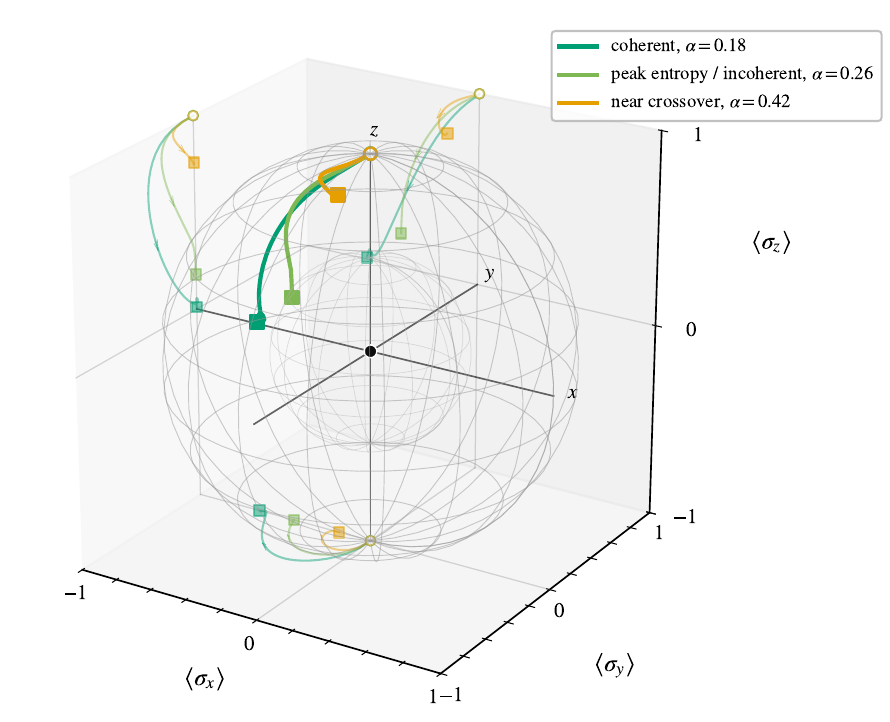}
    \caption{
    Representative three-dimensional Bloch trajectories at $s=0.7$.
    The solid outer wireframe is the pure-state Bloch sphere ($r=1$, $S_{\mathrm{spin}}=0$), and the lighter inner shell marks $r\simeq0.48$, close to the smallest stationary radius among the $s=0.7$ trajectories (maximum $S_{\mathrm{spin}}$).
    The black dot marks the origin ($r=0$), so the distance from this point gives the instantaneous Bloch radius.
    Open circles mark the initial state and squares mark the terminal points.
    }
    \label{fig:bloch_3d}
\end{figure}

This entanglement trend has a simple geometric interpretation in the Bloch-sphere representation.
For a two-level system, the reduced density matrix can be written as
\[
\rho_{\mathrm{spin}} = \frac{1}{2}\bigl(\mathbf{1} + \langle\sigma_x\rangle\,\hat{\sigma}_x + \langle\sigma_y\rangle\,\hat{\sigma}_y + \langle\sigma_z\rangle\,\hat{\sigma}_z\bigr),
\]
with eigenvalues $\lambda_\pm = (1\pm r)/2$, where
\[
r=\sqrt{\langle\sigma_x\rangle^2+\langle\sigma_y\rangle^2+\langle\sigma_z\rangle^2}
\]
is the Bloch radius. The spin entanglement entropy is then
\[
S_{\mathrm{spin}} = -\frac{1+r}{2}\ln\frac{1+r}{2} - \frac{1-r}{2}\ln\frac{1-r}{2},
\]
which is a monotonically decreasing function of $r$ ($S=0$ at $r=1$ and $S=\ln 2$ at $r=0$).
Thus the formation of stationary entanglement corresponds to the trajectory settling onto a fixed-radius shell.
Figure~\ref{fig:bloch_3d} gives three representative three-dimensional trajectories at $s=0.7$: a coherent trajectory ($\alpha=0.18$), a fixed-$s$ peak-entropy trajectory in the incoherent region ($\alpha=0.26$), and a stronger-coupling trajectory near the incoherent--pseudo-coherent crossover ($\alpha=0.42$).
These examples show that the selected trajectories differ not only in their population traces but also in how the Bloch vector approaches its terminal shell. The coherent trajectory winds before settling, the peak-entropy trajectory terminates near the smallest-radius shell, and the stronger-coupling trajectory approaches a larger terminal radius more directly.

\begin{figure}
    \centering
    \includegraphics[width=\linewidth]{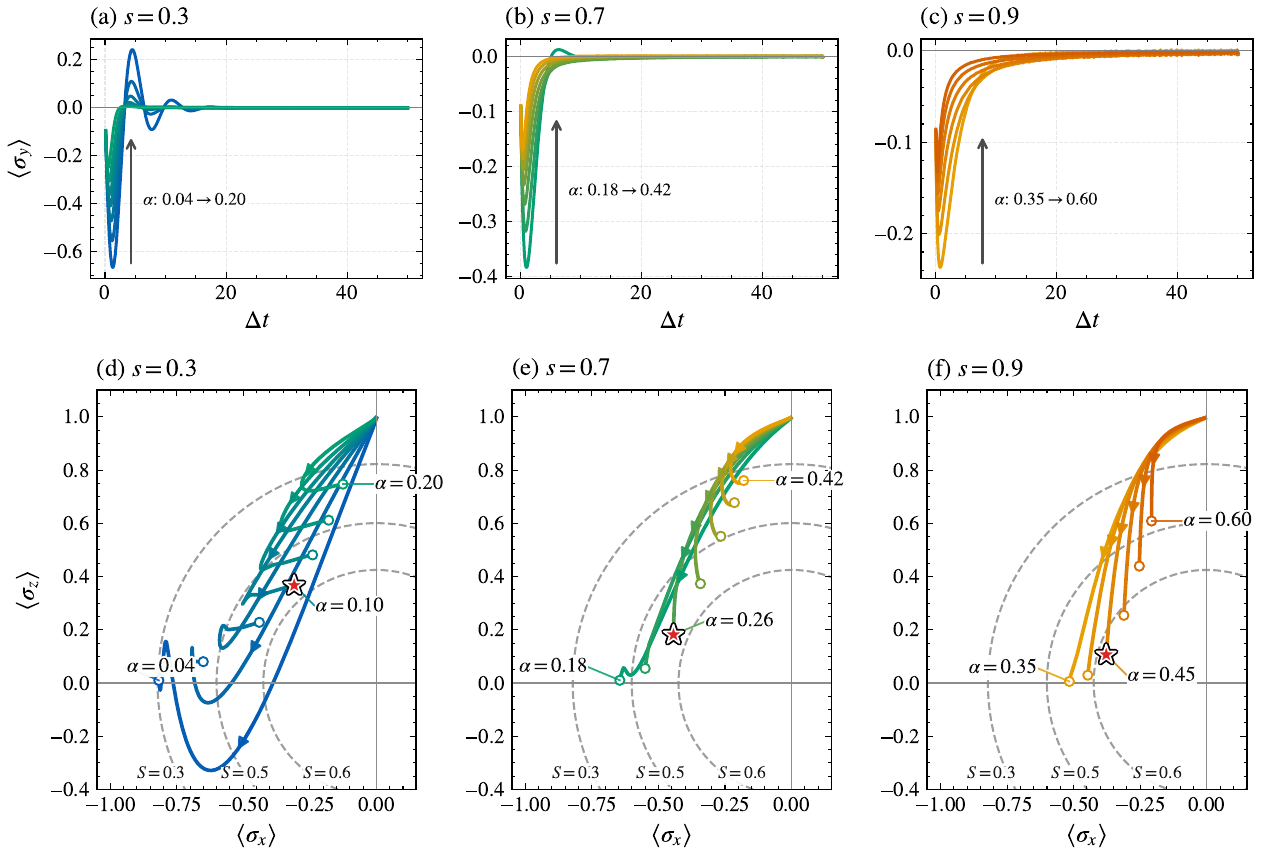}
    \caption{
    Tunneling current and Bloch-sphere trajectories across the dynamical crossover.
    Panels (a)--(c) show the tunneling current $\langle\sigma_y(t)\rangle$ for representative coupling strengths at fixed $s$ ($\alpha=0.04$--$0.20$ for $s=0.3$, $\alpha=0.18$--$0.42$ for $s=0.7$, and $\alpha=0.35$--$0.60$ for $s=0.9$).
    Panels (d)--(f) show the corresponding projected trajectories in the $(\langle\sigma_x\rangle,\langle\sigma_z\rangle)$ plane; dashed circles indicate constant spin entanglement entropy contours.
    The red star marks the fixed-$s$ peak-entropy trajectory.
    Color encodes coupling strength $\alpha$ within each panel, progressing from weak (cool) to strong (warm) coupling.
    }
    \label{fig:bloch_sy}
\end{figure}

To quantify how the trajectories reach their stationary shells, we analyze the tunneling current $\langle\sigma_y(t)\rangle$ together with projected Bloch trajectories in the $(\langle\sigma_x\rangle,\langle\sigma_z\rangle)$ plane (Fig.~\ref{fig:bloch_sy}).
The analyzed coupling strengths span $\alpha=0.04$--$0.20$ for $s=0.3$, $\alpha=0.18$--$0.42$ for $s=0.7$, and $\alpha=0.35$--$0.60$ for $s=0.9$.
The dashed circles in panels (d)--(f) mark constant-entropy contours; trajectories that terminate closer to the origin have higher stationary entanglement.
The decay of $\langle\sigma_y\rangle$ marks the cessation of coherent population transfer; once it vanishes, the Bloch radius stabilizes and $S_{\mathrm{spin}}$ reaches its plateau.
In the unbiased model, $d\langle\sigma_z\rangle/dt = 2\Delta\langle\sigma_y\rangle$, so $\langle\sigma_y\rangle$ represents the instantaneous tunneling current and must vanish at stationarity.
The data confirm this, with $\langle\sigma_y\rangle$ decaying to values below $10^{-3}$ at late times for all parameters studied.
The decay timescale of $\langle\sigma_y\rangle$ (approximately 3 a.u.\ at $s=0.3$, 7 a.u.\ at $s=0.7$, and 19 a.u.\ at $s=0.9$ for the fixed-$s$ peak-entropy trajectories) sets the timescale on which the projected $(\langle\sigma_x\rangle, \langle\sigma_z\rangle)$ trajectory settles to its terminal point.
The $\langle\sigma_x\rangle$ and $\langle\sigma_z\rangle$ components stabilize shortly after $\langle\sigma_y\rangle$ vanishes, confirming that the Bloch vector reaches its stationary configuration once the tunneling current ceases.
An exact consequence of $\langle\sigma_y\rangle\to 0$ is that the stationary Bloch vector lies in the $(\langle\sigma_x\rangle, \langle\sigma_z\rangle)$ plane, so the stationary radius is $r = \sqrt{\langle\sigma_x\rangle^2 + \langle\sigma_z\rangle^2}$ with no out-of-plane component.
The data confirm this to numerical precision ($|r_{\mathrm{full}}-r_{xz}| < 10^{-5}$ for all parameters), validating that the projected trajectories in panels (d)--(f) capture the full stationary state geometry without information loss.
This is precisely why the constant-entropy contours in panels (d)--(f) can be drawn as circles in the $(\langle\sigma_x\rangle, \langle\sigma_z\rangle)$ plane. The stationary entanglement is completely determined by the terminal position in this projection, with the distance from the origin giving the Bloch radius and hence $S_{\mathrm{spin}}$.

The $xz$ projections also provide a geometric reinterpretation of the pseudo-coherent regime. In the population-based classification, pseudo-coherent dynamics is defined by a single minimum in $\langle\sigma_z(t)\rangle$ followed by partial recovery toward localization. The projected Bloch trajectories in panel (d) reveal that this $\langle\sigma_z\rangle$ minimum corresponds to a pronounced turn in the $xz$ plane. The spin first develops a large negative $\langle\sigma_x\rangle$ (reaching $\langle\sigma_x\rangle\approx -0.59$ at $\alpha=0.08$, $s=0.3$), representing a tunneling attempt in which the Bloch vector swings toward the opposite well. The bath then arrests this attempt and the trajectory turns back, with $\langle\sigma_x\rangle$ recovering toward a less negative terminal value while $\langle\sigma_z\rangle$ partially relocalizes. The pseudo-coherent regime thus corresponds geometrically to a failed tunneling excursion that is visible as a hook-shaped trajectory in the Bloch-sphere projection, rather than the simple $z$-axis descent that the population dynamics alone would suggest.
The coherent-regime trajectories, by contrast, execute multiple such excursions (the winding pattern in panels d--f) before settling, consistent with their oscillatory $\langle\sigma_z\rangle$ traces.

The peak-entropy trajectories (stars) correspond to the fixed-$s$ parameter points where the stationary Bloch radius is smallest, representing the maximum spin-bath correlation for each selected $s$ value ($\alpha=0.10$ for $s=0.3$, $\alpha=0.26$ for $s=0.7$, and $\alpha=0.45$ for $s=0.9$).
Quantitatively, the minimum stationary Bloch radius at these fixed-$s$ maxima is $r_{\mathrm{min}}\approx 0.48$ for $s=0.3$ and $s=0.7$, decreasing to $r_{\mathrm{min}}\approx 0.39$ for $s=0.9$.
For comparison, the weakest-coupling trajectories terminate at $r\approx 0.50$--$0.63$ (moderate entanglement) while the strongest-coupling trajectories terminate at $r\approx 0.64$--$0.78$ (lower entanglement, approaching localization).
The non-monotonic dependence of the terminal radius on $\alpha$ directly produces the dome structure in $S_{\mathrm{stable}}$.

\subsection{Mode-Resolved Bath Entanglement Dynamics}
\label{results:ent}

To understand the environmental contribution to the entanglement structure, we analyze the entanglement dynamics of individual bosonic modes.
The mode entanglement entropy is obtained from the one-mode reduced density matrix by tracing out all other degrees of freedom.
We also compute the spin--mode 2-body entropy $S_{\mathrm{spin},j}(t)$ from the reduced density matrix of the spin and an individual bosonic mode.
This quantity measures how the selected spin--mode pair is entangled with the remaining bath degrees of freedom.
For comparison with the one-mode entropy, we use it to form the spin--boson-mode mutual information
$I_{\mathrm{spin}:j}(t)=S_{\mathrm{spin}}(t)+S_j(t)-S_{\mathrm{spin},j}(t)$.

\begin{figure}
    \centering
    \includegraphics[width=\linewidth]{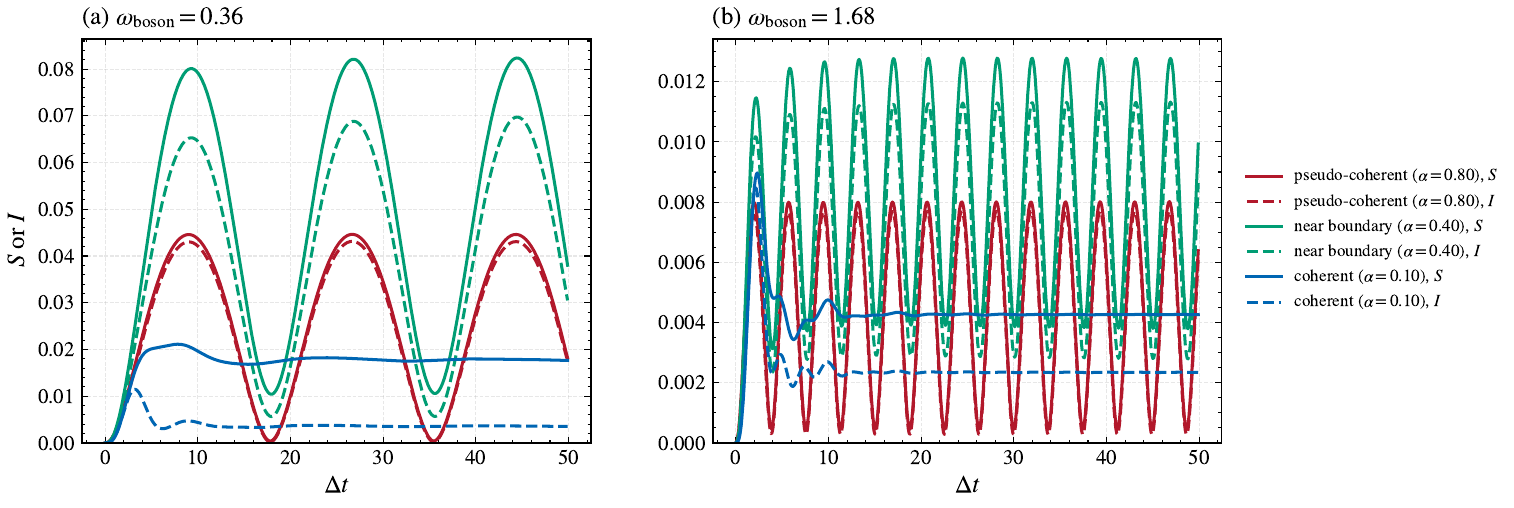}
    \caption{
    Mode-resolved environmental entanglement in representative dynamical regimes at $s=0.7$.
    Panels (a) and (b) show the mode entanglement entropy $S$ (solid) and the spin--boson-mode mutual information $I$ (dashed), computed from the spin--mode 2-body entropy, for a low-frequency mode ($\omega_{\mathrm{boson}}\approx 0.36$) and a higher-frequency mode ($\omega_{\mathrm{boson}}\approx 1.68$), respectively.
    In the pseudo-coherent and near-boundary non-coherent cases, both quantities exhibit periodic oscillations whose magnitudes are close, indicating that the mode is primarily correlated with the spin.
    }
    \label{fig:S_I}
\end{figure}

Figure~\ref{fig:S_I} compares the mode entanglement entropy $S$ and the spin--boson-mode mutual information $I$ for representative points in different dynamical regimes.
In the pseudo-coherent and near-boundary non-coherent cases, both quantities exhibit long-lived periodic oscillations.
Their magnitudes are close ($S \approx I$), which indicates that the mode entanglement is largely accounted for by spin--mode correlations, with weaker mode--mode correlations.
Quantitatively, for the low-frequency mode ($\omega_{\mathrm{boson}}\approx 0.36$) at $s=0.7$, the time-averaged ratio $\langle I\rangle/\langle S\rangle$ is 0.96 in the pseudo-coherent case and 0.81 near the incoherent--pseudo-coherent boundary.
In the coherent regime, by contrast, the mode entropy exceeds the spin--boson-mode mutual information ($S>I$), indicating that coherent spin dynamics mediate additional correlations among bath modes.
The difference $S-I$ therefore provides a mode-level indicator of residual bath-mode correlations, which can reveal organizational differences that are not visible from the scalar stationary spin entropy alone.

To quantify the oscillations, we define the period $T$ as the time interval between successive local minima and the amplitude $S_{\mathrm{Amp}}$ as half the difference between adjacent extrema:
\begin{equation}
T = t_{\mathrm{min},n} - t_{\mathrm{min},n-1}, \qquad
S_{\mathrm{Amp}} = \frac{S_{\mathrm{max}} - S_{\mathrm{min}}}{2}.
\end{equation}
The corresponding oscillation frequency is $\Omega=1/T$.

\begin{figure}
    \centering
    \includegraphics[width=\linewidth]{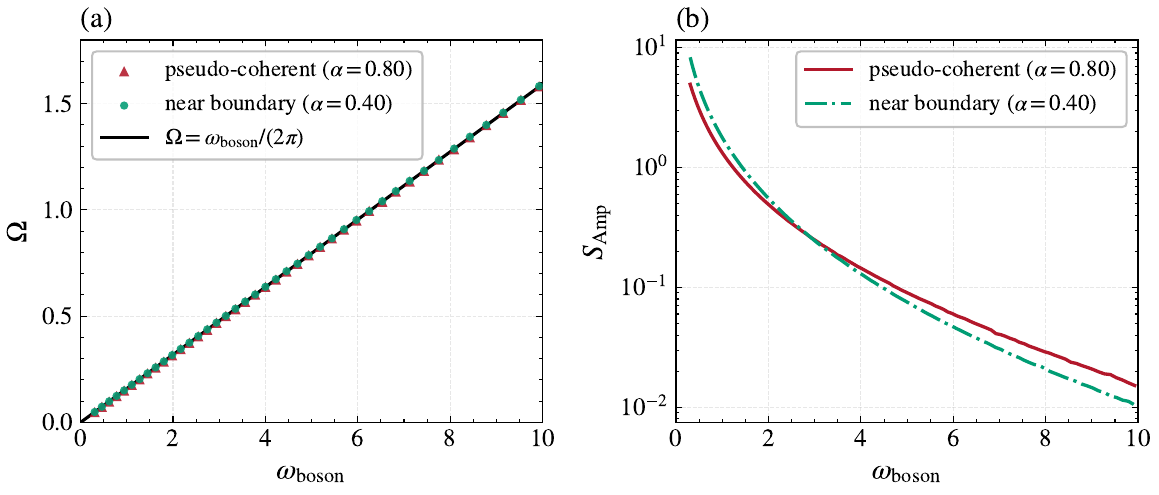}
    \caption{
    Frequency scaling of mode entanglement oscillations at $s=0.7$.
    (a) The entanglement oscillation frequency $\Omega$ scales linearly with the bosonic mode frequency $\omega_{\mathrm{boson}}$, with slope $\approx 1/(2\pi)$.
    (b) The oscillation amplitude $S_{\mathrm{Amp}}$ decays rapidly with increasing $\omega_{\mathrm{boson}}$, showing that low-frequency modes dominate the oscillatory component of environmental entanglement.
    }
    \label{fig:boson_freq}
\end{figure}

As shown in Fig.~\ref{fig:boson_freq}(a), the entanglement oscillation frequency $\Omega$ scales linearly with the bosonic mode frequency $\omega_{\mathrm{boson}}$ in both non-coherent representative cases, with slope approximately $1/(2\pi)$.
This indicates that the entanglement dynamics of each mode is governed by its bare oscillator frequency rather than a polaron-renormalized frequency.
The physical interpretation is that the entanglement oscillation reflects the free precession of each mode's displacement in phase space, while the spin-bath coupling determines only the oscillation amplitude (how far the mode is displaced) rather than the oscillation rate.

The amplitude $S_{\mathrm{Amp}}$ decays rapidly with increasing $\omega_{\mathrm{boson}}$ [Fig.~\ref{fig:boson_freq}(b)].
Modes with frequencies below approximately $4\Delta$ exhibit pronounced entanglement oscillations, whereas higher-frequency modes contribute much more weakly.
This behavior is consistent with the sub-Ohmic spectral density, where low-frequency modes carry enhanced coupling weight and are more strongly displaced by the spin-bath interaction.
The rapid decay indicates that, in these representative $s=0.7$ cases, the stationary spin entanglement entropy $S_{\mathrm{stable}}$ is primarily controlled by the slow bath modes that dominate the dissipative dynamics.

To further verify this conclusion, we compute the density-weighted time-averaged mode entanglement $S_{\mathrm{mean}}(\omega_{\mathrm{boson}}) = \rho(\omega_{\mathrm{boson}}) N^{-1} \sum_{n=1}^{N} S_n(\omega_{\mathrm{boson}})$, which also decays rapidly from low to high frequencies.
Together with the frequency-resolved analysis, these results provide a consistent picture for the representative trajectories studied here. The stationary spin-bath entanglement is mainly controlled by low-frequency bath modes whose strong coupling and slow dynamics set the entanglement scale, while the residual $S-I$ signal distinguishes how bath-mode correlations are organized across dynamical regimes.

\section{Conclusions}
\label{conclusion}

We have investigated spin-bath entanglement across the dynamical phases of the sub-Ohmic spin-boson model using TTN-TDVP-PS, covering the parameter space $s=0.1$--$0.9$ and $\alpha=0.01$--$1.0$.

The population-based dynamical phase diagram, constructed from $\langle\sigma_z(t)\rangle$ trajectories and classified into coherent, incoherent, and pseudo-coherent regimes, agrees well with previous QUAPI results. Beyond this classification, the spin entanglement entropy provides a complementary perspective on the phase structure. The stationary spin entanglement entropy $S_\mathrm{stable}$ forms a dome-shaped landscape in the $(s,\alpha)$ plane whose ridge traces the parameters of large spin-bath correlation. At small $s$, this ridge broadly follows the population-based phase boundary. At large $s$, where the population dynamics branches into separate coherent-incoherent and incoherent-pseudo-coherent boundaries, the scalar entropy landscape does not reproduce this branching. Its ridge remains single-valued and lies within the incoherent region, indicating that the stationary entanglement maximum does not separately track both population-based transitions.

The Bloch-sphere analysis clarifies the geometric content of this observation. The spin entanglement entropy is a monotonic function of the Bloch radius, so the entropy plateau corresponds to the trajectory settling onto a fixed-radius shell. The tunneling current $\langle\sigma_y(t)\rangle$ controls the approach to this shell. Once it decays, the Bloch radius stabilizes and the entropy reaches its stationary value. At fixed-$s$ peak-entropy parameters, the stationary Bloch radius is smallest and the spin is maximally mixed with respect to the bath. The three dynamical regimes are geometrically distinguished by their trajectory morphology on the Bloch sphere (spiral, monotonic, and hook-shaped failed tunneling excursion), with the pseudo-coherent hook providing a geometric reinterpretation of what the population dynamics records as a single $\langle\sigma_z\rangle$ minimum.

Mode-resolved analysis reveals that the environmental entanglement is concentrated in low-frequency bath modes. The entanglement of individual modes oscillates at the bare boson frequency rather than a renormalized frequency, with amplitudes that decay rapidly with mode frequency. The spin--boson-mode mutual information, computed from the spin--mode 2-body entropy, further separates direct spin--mode correlations from residual bath-mode correlations. In the representative coherent case, $S>I$ indicates a larger bath-mode correlation component than in the non-coherent cases where $S\approx I$.

The present analysis can be extended to finite-temperature initial states, biased systems, and structured spectral densities relevant to molecular environments.

\begin{acknowledgement}
The authors thank Zhigang Shuai and Shixing Zhang for helpful discussions.
This work was supported by the Shenzhen Science and Technology Program (No. KQTD20240729102028011), the National Natural Science Foundation of China (22573088, 22595405), the Guangdong Basic and Applied Basic Research Foundation (2026A1515010545), the Guangdong Provincial Quantum Science Strategic Initiative (Grant No. GDZX2503001), and the Guangdong Basic Research Center of Excellence for Aggregate Science.
The computation hardware resource is supported by the Open Source Supercomputing Center of S-A-I and the High-Performance Computing Portal, which is under the administration of the Information Technology Services Office (ITSO) at the Chinese University of Hong Kong, Shenzhen.
\end{acknowledgement}

\section*{Data Availability Statement}
The source data will be made available on GitHub.


\bibliography{main}

\end{document}